\documentclass[aps,prl,reprint,amsmath,amssymb,twocolumn]{revtex4} 
\usepackage[british]{babel}
\usepackage{graphicx}
\usepackage{amsmath}

\usepackage{bbm}
\usepackage{amssymb}
\usepackage{varioref}
\usepackage{hyperref}
\usepackage{bibentry}

\usepackage{color}
\usepackage{hyperref}
\usepackage{placeins}

\def\eq{\begin{eqnarray}}
\def\en{\end{eqnarray}}

\begin{document}
\sloppy
\title{Comment on ``Contextuality in Bosonic Bunching''}

\author{Malte C.~Tichy and Christian Kraglund Andersen}
\affiliation{Department of Physics and Astronomy, Aarhus University, DK-8000 Aarhus, Denmark}

\begin{abstract}
Bosonic bunching occurs within quantum physics and can be mimicked classically by noncontextual hidden-variable models, which  excludes this phenomenon as a means to prove stronger-than-quantum contextuality. 
\end{abstract}
\date{\today}
\maketitle 
 Kurzy{\'n}ski {\it et al.}~proposed a system that allegedly  violates the exclusivity principle  \cite{kurzynski}. As illustrated in Fig.~1 in Ref.~\cite{kurzynski}, three bosons are pairwisely combined at a beam splitter at which they bunch and leave in the same random output mode. The events $\underline{a}b$, $\underline{b}c$ and $a\underline{c}$ [in which $\underline{k}$ ($k$) denotes particle $k$ being reflected (transmitted)] are conjectured to be pairwisely exclusive, since the reflection of $a$ excludes its transmission. All event probabilities being $1/2$ due to  bunching, the sum 3/2 violates the exclusivity bound of unity and saturates the  bound allowed by  no-disturbance. However,   following the authors' reasoning, the three events are not exclusive, which disproves the letter's conclusions. 

The attribution of exclusivity to the three events relies on assumption (ii): ``The scattering properties of each boson on the BS [beam splitter] do not depend on which other fiber is connected to the other BS's input port and on the choice of the BS's input port.'' (all  quotes taken from \cite{kurzynski}). 
 Essentially, (ii) states that particles must propagate independently. Such strong assumption is unreasonable to impose on physical theories, since it excludes any interaction. It is violated already by classical systems, and by particles described by quantum mechanics that interact either via a potential  or by an effective  bosonic (fermionic) exchange interaction. 

Despite assumption (ii) being clearly violated by correlated bosons, (ii) is nevertheless upheld in \cite{kurzynski} to corroborate exclusivity: Although the authors realise that ``[...] at least one of the assumptions (i), (ii), and (iii) does not hold'', they do not give up any of them, and conclude: ``there are events that [...] can be considered as exclusive if one takes into account assumptions (i), (ii), and (iii).'' Thus, the exclusivity of the three events -- the letter's very leitmotiv -- is based on the disproved assumption (ii). 
 The formal violation of the exclusivity principle is then neither surprising, nor is it characteristic to identical particles or quantum physics.

As noticed by the  authors, although assumption (ii) is baptized ``noncontextuality'', it is unrelated to   accepted ``traditional'' noncontextuality \cite{cabello}. The traditional definition is referred to in 
  abstract, introduction and conclusion, even though a traditionally noncontextual hidden-variable model reproduces all phenomena    in \cite{kurzynski}: Each particle $j$ ($j=1\dots 3$) is assigned a random 
    hidden variable $0<\lambda_j<1$, two particles $j$, $k$ that impinge on a beam splitter are ejected through output mode $1$ if $\lambda_j  > \lambda_k$ [output is (2,0)], or  mode $2$ if $\lambda_j < \lambda_k$ [output is (0,2)]. The initially, randomly and independently chosen hidden variables \emph{fully} \emph{pre}-determine the outcomes of all possible measurements, which contradicts the dictum that ``it is not possible to assign properties to individual bosons independently of this choice [of measurement setting]'' (which clearly refers to \emph{traditional} contextuality). 
The model perfectly mimics two-boson bunching and reproduces the violation of the exclusivity bound and of the KCBS inequality put forward in \cite{kurzynski}. Despite the authors' claim that no mechanism for such behavior exists which keeps the beam splitter as a ``deterministic memoryless device'' without ``intrinsic randomness'' and ``whose action only depends on values of the variables assigned to individual particles'', our model achieves precisely that.

The saturation of the no-disturbance bound 3/2 in \cite{kurzynski} is accidental: In a modified model, both particles exit through the first mode if $\lambda_j + \delta > \lambda_k$, the average  sum of probabilities becomes $3/2+\delta-\delta^2/2>3/2$. This violation occurs because assumption (1.) in \cite{kurzynski} (\emph{Complementarity}), on which the upper bound of 3/2 relies, is not fulfilled either: The assumption that \emph{two} events out of $\{ \underline{a}b, \underline{b}c,  a \underline{c} \}$ are tested \emph{simultaneously} by a single measurement is based on the violated assumption (ii). By testing  $a$ and $b$, we \emph{cannot} infer how $a$ and $c$ would have behaved. 

Furthermore, the authors claim that ``a system of bosonic particles and a set of measurement events'' is ``capable'' to do something that cannot be done ``using standard quantum events described by projectors''. However, the events in the setup
  are   described by three \emph{nonorthogonal, noncommuting}  projectors. 
Within quantum physics,  the measurements fail to fulfil the  requirements of exclusivity and compatibility assumed 
for noncontextual inequalities or the exclusivity principle \cite{cabello}. We doubt that the indistinguishability of identical particles can add  new features to quantum contextuality: 
Exclusivity and compatibility are \emph{independent} of the \emph{implementation} of a quantum system by one or several, distinguishable or identical particles. In contrast to  identical-particle entanglement, the lack of a well-defined tensor-product structure is unproblematic for   (non)contextuality \cite{cabello2}: The Fock space of $N$ bosons or fermions in $m$ modes is a finite-dimensional Hilbert pace, equivalent to the Hilbert-space of a single particle. Fundamentally speaking, a 
   violation of a theorem that is proven to apply to quantum mechanics is tautologically excluded for any quantum-mechanical  system, be it realized by distinguishable particles, bosons or fermions. 

\emph{Acknowledgements.} The authors thank Klaus M{\o}lmer and David Petrosyan for valuable comments on the manuscript, and Pawe{\l} Kurzy{\'n}ski, Akihito Soeda, Jayne Thompson and Dagomir Kaszlikowski for helpful discussions. M.C.T. and C.K.A. acknowledge funding by the Danish Council for Independent Research and the EU 7th Framework Programme collaborative project iQIT, respectively; both authors acknowledge funding by the Villum Foundation.

\end{document}